\documentclass[aps,prb,%
twocolumn,%
amssymb,amsmath,%
showpacs,showkeys,%
tightenings,%
ejectum,%
breveted]{revtex4}
\usepackage[final]{graphicx}
\usepackage{dcolumn}
\begin{document}
\title{Systematic speedup of path integrals of a generic $N$-fold discretized theory}
\author{A. Bogojevi\'c}\altaffiliation{E-mail: alex@phy.bg.ac.yu}
\author{A. Bala\v{z}}\altaffiliation{E-mail: antun@phy.bg.ac.yu}
\author{A. Beli\'c}\altaffiliation{E-mail: abelic@phy.bg.ac.yu}
\affiliation{Institute of Physics, P.O. Box 57, 11001
Belgrade, Serbia and Montenegro}
\begin{abstract}
We present and discuss a detailed derivation of a new analytical method that systematically improves the convergence of path integrals of a generic $N$-fold discretized theory. We develop an explicit procedure for calculating a set of effective actions $S^{(p)}$, for $p=1,2,3,\ldots$ which have the property that they lead to the same continuum amplitudes as the starting action, but that converge to that continuum limit ever faster. Discretized amplitudes calculated using the $p$ level effective action differ from the continuum limit by a term of order $1/N^p$. We obtain explicit expressions for the effective actions for levels $p\le 9$. We end by analyzing the speedup of Monte Carlo simulations of two different models: an anharmonic oscillator with quartic coupling and a particle in a modified P\"oschl-Teller potential. 
\end{abstract}
\preprint{SCL preprint}
\maketitle

\section{\label{sec:intro} Introduction}

In the functional formalism \cite{feynmanhibbs,feynman} the general quantum mechanical amplitude $A(a,b;T)=\langle b|e^{-T\hat H}|a\rangle$ is given 
in terms of a path integral which is simply the $N\to\infty$ limit of the expression
\begin{equation}
\label{amplitudeN}
A_N(a,b;T)=\left(\frac{1}{2\pi\epsilon_N}\right)^{\frac{N}{2}}\int dq_1\cdots dq_{N-1}\,e^{-S_N}\ .
\end{equation}
The euclidean time interval $[0,T]$ has been subdivided into $N$ equal time steps of length $\epsilon_N=T/N$, with $q_0=a$ and $q_N=b$. $S_N$ is the naively discretized action of the theory. In this paper we will look at theories with action of the form 
\begin{equation}
\label{action}
S=\int_0^Tdt\,\left(\frac{1}{2}\, \dot q^2+V(q)\right)\ .
\end{equation}
Note that we use units in which $\hbar$ and particle mass have been set to unity.
The naively discretized action is in this case simply
\begin{equation}
\label{actionN}
S_N=\sum_{n=0}^{N-1}\left(\frac{\delta_n^2}{2\epsilon_N}+\epsilon_NV(\bar q_n)\right)\ ,
\end{equation}
where $\delta_n=q_{n+1}-q_n$, and $\bar q_n=\frac{1}{2}(q_{n+1}+q_n)$. Key investigations regarding numerical evaluation of path integrals were presented in the reviews of Barker and Henderson \cite{barkerhenderson}, Kalos and Whitlock \cite{kaloswhitlock}, and Ceperley \cite{ceperley}, as well as in the papers by Pollock and Ceperley \cite{pollockceperley} and Barker \cite{barker}. A modern and extensive reference on the subject of path integrals is given in the latest edition of the textbook by Kleinert\cite{kleinert}.

As we can see, the very definition of path integrals makes it necessary to make the transition from the continuum to the discretized theory. This discretization, however, is far from unique. In fact, the details of the discretization procedure are extremely important both for analytical and numerical
treatment of path integrals. This dependence on discretization procedure has been one of the principle impediments to creating a consistent mathematical theory of path integration. On the numerical side, this manifests itself in the fact that path integral simulations remain notoriously demanding of computing time -- so much so that certain path integral calculations serve as benchmarks for new generations of supercomputers.

For this reason let us note that we have the freedom to make two important choices that will not affect the final result, i.e. the continuum amplitude we seek to calculate. First, we have the freedom to choose the point in the interval $[q_n,q_{n+1}]$ in which to evaluate the potential $V$. It is well known that different points correspond to different ordering prescriptions in the operator formalism. The choice of the middle point $\bar q_n$ is the most common one. It corresponds to the symmetric or Weyl ordering of operators $\hat p$ and $\hat q$, and so always leads to a hermitean expression for the hamiltonian $\hat H$. Two other prescriptions are also often used. The left ordering prescription evaluates the potential at $q_n$, the left boundary of the above interval (in the operator formalism this corresponds to taking the $\hat p$'s to the left of the $\hat q$'s in all the products that appear in the hamiltonian). Similarly, one defines the right ordering prescription. Although they lead to somewhat simpler looking expressions, the left and right prescriptions do not in general give hermitean hamiltonians. Let us note, however, that the class of theories given by eq.~(\ref{action}) has a hamiltonian that is the sum of a $\hat p$-dependent kinetic term and a $\hat q$-dependant potential term and so has no ordering ambiguities. In this case different prescriptions lead to the same continuum amplitude -- the discretized amplitudes do differ, but they tend to the same continuum limit. 

The second, and more important freedom related to our choice of discretized action has to do with the freedom to introduce additional terms that explicitly vanish in the continuum limit. We will designate such discrete actions as effective actions. For example, the term
\begin{equation}
\sum_{n=0}^{N-1}\epsilon_N\,\delta_n^2\,g(\bar q_n)\ ,
\end{equation} 
where $g$ is regular when $\epsilon_N\to0$, does not change the continuum physics since it goes over into $\epsilon_N^2\int^T_0dt\,\dot q^2\,g(q)$, i.e. it vanishes as $\epsilon_N^2$. Although such additional terms do not change the continuum physics they do affect the speed of convergence to that continuum limit. 

The aim of this paper is to give a detailed exposition of a systematic analysis that leads to the best solution in the class of all equivalent effective actions, e.g. the effective action that leads to the fastest convergence of the associated discrete amplitude $A_N$ to the continuum expression. The uncovered speedup in the path integral algorithm is a direct consequence of new analytical input that has come from the study of the relation between discretizations of differing coarseness of the same theory. We have given a brief presentation of these ideas in a recent paper \cite{prl}.

The calculations that will be presented turn out to be simplest in the mid-point prescription. Before we proceed with them it is useful to spend the next section in a brief overview of known results dealing with the speed of convergence to the continuum limit. 

\section{\label{sec:overview} Brief overview of known results}

In this section we will compare the speed of convergence of several different prescriptions to the continuum limit. The naive mid-point prescription satisfies
\begin{equation}
\label{mid}
A_N(a,b;T)^{mid}=A(a,b;T)+O(1/N)\ ,
\end{equation}
for all $a$ and $b$. By naive we mean that we use the naively discretized action given in eq.~(\ref{actionN}). On the other hand, in the naive left prescription the amplitude for $a\to a$ converges much faster
\begin{equation}
\label{leftdiagonal}
A_N(a,a;T)^{left}=A(a,a;T)+O(1/N^2)\ .
\end{equation}
This behavior can be easily shown both analytically and numerically. We note in passing that this is strongly related to the well known result for the partition function evaluated using naive discretization in the left prescription \cite{krajewskimuser}, which follows directly from the above amplitude by integrating over $a$ (to get the trace) and writing the time of propagation $T$ as the inverse temperature $\beta$, 
\begin{equation}
\label{partition}
Z_N(\beta)^{left}=Z(\beta)+O(1/N^2)\ .
\end{equation}
However, going back to the language of amplitudes, it is also easy to show that the amplitudes for different initial and final states converge slower, i.e. for $a\ne b$ we have
\begin{equation}
\label{leftoffdiagonal}
A_N(a,b;T)^{left}=A(a,b;T)+O(1/N)\ .
\end{equation}

The problems with the speed of convergence of off-diagonal amplitudes in the left prescription can be fixed very easily. We find that for all $a$ and $b$ we have
\begin{eqnarray}
\label{leftright}
\lefteqn{\frac{1+e^{\epsilon_N(V(a)-V(b))}}{2}\,A_N(a,b;T)^{left}=}\nonumber\\
&&{}\qquad\qquad\qquad\qquad=A(a,b;T)+O(1/N^2)\ .
\end{eqnarray}

Although not related to the central investigation in this paper, let us present a proof of eq.~(\ref{leftright}) as an illustration of the use of the Trotter formula 
\begin{equation}
\label{trotter}
e^{\hat A +\hat B}=\lim_{N\to\infty} \left(e^{\hat A/N}e^{\hat B/N}\right)^N\ .
\end{equation}
Using it we can easily show the validity of the formal expression $A_{-N}(a,b;T)^{left}= A_{N}(a,b;T)^{right}$. On the other hand, from eq.~(\ref{amplitudeN}) we find that $A_{N}(a,b;T)^{right}=e^{\epsilon_N(V(a)-V(b))}A_{N}(a,b;T)^{left}$. As a result we see that the left hand side of eq.~(\ref{leftright}) is simply the average of $A_N$ and $A_{-N}$ and so has an expansion in even powers of $1/N$. 

We end this section by presenting the result obtained by Takahashi and Imada \cite{takahashiimada} and independently by Li and Broughton \cite{libroughton}. In these papers the authors used a generalized form \cite{deraedt2} of the Trotter formula to increase the speed of convergence of the discretized partition function. Their final result is a derivation of a formula for the effective potential $V^{eff}=V+\frac{1}{24}\epsilon_N^2(V')^2$. The authors showed that by using this effective potential (in the left prescription) one gets 
\begin{equation}
\label{eff}
Z_N(\beta)^{eff}=Z(\beta)+O(1/N^4)\ .
\end{equation}

A recent analysis of this method can be found in Jang et al \cite{jangetal}. Let us note that the crucial step in the derivation of the above effective potential from the generalized Trotter formula uses the cyclic property of the trace, i.e. the above increase in the speed of convergence only holds for the partition function and not the amplitudes. A direct numerical simulation shows that amplitudes calculated using this effective potential converge just as fast as the amplitudes in the naive left prescription. Said another way, it is only the integral over all the diagonal amplitudes that has the $O(1/N^4)$ behavior and not any individual amplitude. A recent investigation by Bond et al. \cite{bondetal} has uncovered a $O(1/N^6)$ behavior, however, not for the case of a generic theory. At the end let us mention that several related investigations dealing with speed of convergence have focused on improvements in short time propagation \cite{makrimiller1,makrimiller2,makri} or the action \cite{alfordetal}.  

\section{\label{sec:discretizationhalving} Relation between different discretizations}

The aim of this and the following section is to present a systematic exposition of the relation between different discretizations of the same path integral. Throughout we will work in the mid-point prescription. We start by studying the relation between the $2N$-fold and $N$-fold discretizations of a given amplitude. From eq.~(\ref{amplitudeN}) we see that we can write the $2N$-fold amplitude as
\begin{equation}
\label{amplitude2N}
A_{2N}(a,b;T)=\left(\frac{1}{2\pi\epsilon_N}\right)^{\frac{N}{2}}\int dq_1\cdots dq_{N-1}\,e^{-\widetilde S_N}\ ,
\end{equation}
i.e. in the form of an $N$-fold amplitude given in terms of a new action $\widetilde S_N$ determined by
\begin{equation}
\label{tildeS}
e^{-\widetilde S_N}=\left(\frac{2}{\pi\epsilon_N}\right)^{\frac{N}{2}}\int dx_1\cdots dx_N\,e^{- S_{2N}}\ ,
\end{equation}
where $S_{2N}$ is nothing but the $2N$-fold discretization of the starting action. In the above formulas we have, for convenience, written the $2N$-fold discretized coordinates $Q_0,Q_1,\ldots,Q_{2N}$ in terms of $q$'s and $x$'s in the following way: $Q_{2k}=q_k$ and $Q_{2k-1}=x_k$. Note that we have $q_0=a$, $q_N=b$, while the $N-1$ remaining $q$'s play the role of the dynamical coordinates in the $N$-fold discretized theory. The $x$'s are the $N$ remaining intermediate points that we integrate over in eq.~(\ref{tildeS}). 

We wish to have $\widetilde S_N$ belong to the same class of actions as $S_N$. It is not difficult to show that the naively discretized action does not satisfy this requirement, i.e. the integration of eq.~(\ref{tildeS}) will yield new types of terms in $\widetilde S_N$. In fact, the class of actions closed to transformation (\ref{tildeS}) is of the form
\begin{eqnarray}
\label{actions}
\lefteqn{
S_N=\sum_{n=0}^{N-1}\bigg(\frac{\delta_n^2}{2\epsilon_N}+\epsilon_N\,V(\bar q_n)+\epsilon_N\,\delta_n^2\,g_1(\bar q_n)\,+}\nonumber\\
&&{}\quad+\,\epsilon_N\,\delta_n^4\,g_2(\bar q_n)+\epsilon_N\,\delta_n^6\,g_3(\bar q_n)+\ldots\bigg)\ .
\end{eqnarray}
The functions appearing in the above effective action also depend on the time step $\epsilon_N$. We choose not to display this dependence explicitly in order to have a more compact notation. What is important is that all of these functions are regular in the $\epsilon_N\to 0$ limit. Note that these effective actions are equivalent to our starting action, i.e. they all have the same continuum amplitudes as the starting theory. Using eq.~(\ref{tildeS}) and (\ref{actions}) one can easily derive the following integral relation which determines the functions $\widetilde V, \widetilde g_1, \widetilde g_2, \ldots$ in the new action $\widetilde S$ in terms of the related functions in the starting action:
\begin{widetext}
\begin{equation}
\label{integral}
\exp\bigg(-\epsilon_N\Big(\widetilde V(\bar q_n)+\delta_n^2\,\widetilde g_1(\bar q_n)+\delta_n^4\,\widetilde g_2(\bar q_n)+\ldots\Big)\bigg)=
\left(\frac{2}{\pi\epsilon_N}\right)^{\frac{1}{2}}\int_{-\infty}^{+\infty} 
dy\,\exp\left(-\frac{2}{\epsilon_N}y^2\right)F(\bar q_n+y)\ ,
\end{equation}
where
\begin{eqnarray}
\label{F}
\lefteqn{
-\frac{1}{\epsilon_N}\ln F(x)=\frac{1}{2}\,V\left(\frac{q_{n+1}+x}{2}\right)+\frac{1}{2}\,V\left(\frac{x+q_n}{2}\right)+}\nonumber\\
&&{}\qquad\qquad\qquad+\frac{1}{2}(q_{n+1}-x)^2\,g_1\left(\frac{q_{n+1}+x}{2}\right)+\frac{1}{2}(x-q_n)^2\,g_1\left(\frac{x+q_n}{2}\right)+\nonumber\\
&&{}\qquad\qquad\qquad+\frac{1}{2}(q_{n+1}-x)^4\,g_2\left(\frac{q_{n+1}+x}{2}\right)+\frac{1}{2}(x-q_n)^4\,g_2\left(\frac{x+q_n}{2}\right)+\ldots\ .
\end{eqnarray}
\end{widetext}

The above integral equation can be solved for the simple cases of a free particle and a harmonic oscillator, and gives the well known results. Let us note that the integral in eq.~(\ref{integral}) is in a form that is ideal for an asymptotic expansion \cite{erdelyi}, whatever the potential. The time step $\epsilon_N$ plays the role of a small parameter (in complete parallel to the role $\hbar$ plays in the usual semi-classical, or loop, expansion of quantum theories). As is usual, the above asymptotic expansion is carried through by first Taylor expanding $F(\bar q + y)$ around $\bar q$ and then by doing the remaining Gaussian integrals. We find
\begin{eqnarray}
\label{expansion}
\lefteqn{\left(\frac{2}{\pi\epsilon_N}\right)^{\frac{1}{2}}\int_{-\infty}^{+\infty} 
dy\,\exp\left(-\frac{2}{\epsilon_N}y^2\right)F(\bar q_n+y)=}\nonumber\\
&&{}\qquad\qquad =\sum_{m=0}^{\infty}\frac{F^{(2m)}(\bar q_n)}{(2m)!!}\,\left(\frac{\epsilon_N}{4}\right)^m\ ,
\end{eqnarray}
where we have assumed that $\epsilon_N<1$, i.e. that $N>T$. 

Finally, from eq.~(\ref{integral}) and (\ref{expansion}) we get
\begin{eqnarray}
\label{formula}
\lefteqn{\widetilde V(\bar q_n)+\delta_n^2\,\widetilde g_1(\bar q_n)+\delta_n^4\,\widetilde g_2(\bar q_n)+\ldots=}\nonumber\\
&&=-\frac{1}{\epsilon_N}\ln\left[\sum_{m=0}^{\infty}\frac{F^{(2m)}(\bar q_n)}{(2m)!!}\,\left(\frac{\epsilon_N}{4}\right)^m\right]\ .
\end{eqnarray}
All that remains is to calculate the $F^{(2m)}(\bar q_n)$'s using eq.~(\ref{F}) and to expand the potential and all the functions $g_k$ 
around the mid-point $\bar q_n$. For this second step we make use of the simple relations that follow from the definitions of $\bar q_n$ and $\delta_n$:
$q_{n+1}-\bar q_n =  \bar q_n - q_n = \delta_n/2$, $q_{n+1}+\bar q_n = 2\bar q_n + \delta_n/2$, and $\bar q_n+q_n = 2\bar q_n - \delta_n/2$.
Using these relations we can expand a typical term like $(q_{n+1}-\bar q_n)\,g_1'\left(\frac{q_{n+1}+\bar q_n}{2}\right)$ to obtain
$$
\frac{\delta_n}{2}\,g_1'\left(\bar q_n+\frac{\delta_n}{4}\right)=
\frac{\delta_n}{2}\,g_1'(\bar q_n)+\frac{\delta_n^2}{8}\,g_1''(\bar q_n)+\ldots\ .
$$

For example, the expansion of $\widetilde S$ up to $\epsilon_N^3$ is a rather simple exercise. We find the following functional relations
\begin{eqnarray}
\label{halvingrelations}
\widetilde V &=&
V+\epsilon_N\left[{\frac{1}{4}}g_1+\frac{1}{32}V''\right]+\nonumber\\
&&+\epsilon_N^2\,\left[\frac{3}{16}g_2-\frac{1}{32}V'\,^2+\frac{1}{2048}V^{(4)}+\frac{3}{128}g_1''\right]\nonumber\\
\widetilde g_1 &=& \frac{1}{4}g_1+\frac{1}{32}V''+\\
&&+\epsilon_N\left[\frac{3}{8}g_2+\frac{1}{1024}V^{(4)}-\frac{1}{64}g_1''\right]\nonumber\\
\widetilde g_2 &=& \frac{1}{16}g_2+\frac{1}{6144}V^{(4)}+\frac{1}{128}g_1''\ .\nonumber
\end{eqnarray}

Note that in the above relations we expanded $\widetilde V$ up to $\epsilon_N^2$, $\widetilde g_1$ up to $\epsilon_N$ and $\widetilde g_2$ up to $\epsilon_N^0$. We also disregarded all the higher $\widetilde g_k$'s. The reason for this is that the short time propagation of a generic theory satisfies $\delta_n^2\propto\epsilon_N$ while the $g_k$ term enters the action multiplied by $\delta_n^{2k}$. In general, if we expand the new action $\widetilde S$ to $\epsilon_N^p$ we need to evaluate only $\widetilde V$ (up to $\epsilon_N^{p-1}$) and the first $p-1$ functions $\widetilde g_k$ (up to $\epsilon_N^{p-1-k}$). Although straight-forward, the task of calculating $\widetilde S$ to large powers of $\epsilon_N$ is quite tedious; using the symbolic algebra package MATHEMATICA 5.0 we have analytically solved the corresponding expressions up to $p\le 9$. The memory requirements for this calculation grow exponentially with $p$: the $p=9$ calculation used just under 2 GB of computer memory. 

At this point it is important to comment on what has been achieved so far. Evaluating $\widetilde S$ to $\epsilon_N^p$ and using this new action to calculate the $N$-fold discretized amplitude $\widetilde A_N$ we find
\begin{equation}
\label{halving}
\widetilde A_N(a,b;T) = A_{2N}(a,b;T)+O(\epsilon_N^p)\ ,
\end{equation}
so that, up to $O(\epsilon_N^p)$, this amplitude is the same as the $2N$-fold amplitude calculated with our starting action. In this way we have halved the discretization from $2N$ to $N$. Therefore, a coarser $N$-fold discretization using $\widetilde S_N$ does the same job as the $2N$-fold discretization of the starting theory. In the next section we will consider the iteration of this halving procedure. We will derive and solve the recursive relations that connect up the $2^s N$-fold and $N$-fold discretizations. In particular, we will focus on the continuum limit solution when $s\to\infty$, i.e. the solution that connects up the continuum theory with its $N$-fold discretization. 

\section{\label{sec:recursion} Recursive halving}

The iterative process of halving that starts from the $2^s N$-fold discretization is governed by a recursive relation. From the $p=3$ case given in 
eq.~(\ref{halvingrelations}), used in the previous section to illustrate the general discretization halving scheme, we directly get the sought-after $p=3$ level system of recursive relations
\begin{widetext}
\begin{eqnarray}
\label{rec:p3}
V_{k+1} &=&
V_{k}+\frac{\epsilon_N}{2^{s-k-1}}\left[{\frac{1}{4}}(g_1)_{k}+\frac{1}{32}V_{k}''\right]+
\frac{\epsilon_N^2}{2^{2(s-k-1)}}\,\left[\frac{3}{16}(g_2)_{k}-\frac{1}{32}V_{k}'\,^2+\frac{1}{2048}V_{k}^{(4)}+\frac{3}{128}(g_1)_{k}''\right]\nonumber\\
(g_1)_{k+1} &=& \frac{1}{4}(g_1)_{k}+\frac{1}{32}V_{k}''+
\frac{\epsilon_N}{2^{s-k-1}}\left[\frac{3}{8}(g_2)_{k}+\frac{1}{1024}V_{k}^{(4)}-\frac{1}{64}(g_1)_{k}''\right]\\
(g_2)_{k+1} &=& \frac{1}{16}(g_2)_{k}+\frac{1}{6144}V_{k}^{(4)}+\frac{1}{128}(g_1)_{k}''\ .\nonumber
\end{eqnarray}
\end{widetext}
In the above relations $k=0,1,2,\ldots,s-1$. The zeroth iterate corresponds to the starting action, the last iterate to the effective action that gives an equivalent $N$-fold discretization. The $\epsilon_N/2^{s-k-1}$ terms represent the time step of the $k$-th iterate in the discretization halving procedure. 

Although the above system of recursive relations is non-linear it is in fact quite straight-forward to solve if we remember that the system itself was derived via an expansion in $\epsilon_N$. Having this in mind we first write all the functions as expansions in powers of $\epsilon_N$ that are appropriate to the level $p$ we are working at. In this case we are illustrating the procedure for $p=3$, so we have   
\begin{eqnarray}
\label{expand}
V_k &=& A_k + \frac{\epsilon_N}{2^{s-k-1}}B_k+\left(\frac{\epsilon_N}{2^{s-k-1}}\right)^2C_k\nonumber\\
(g_1)_k &=& D_k + \frac{\epsilon_N}{2^{s-k-1}}E_k\\
(g_2)_k &=& F_k\ .\nonumber
\end{eqnarray}
Putting this into the $p=3$ level system of recursive relations given in eq.~(\ref{rec:p3}) we find that $A_{k+1}=A_k$. Using the initial condition $A_0=V$ we find that $A_k=V$ for all $k$. Using this the remaining equations form the following set of linear recursive relations
\begin{eqnarray}
\label{linear}
2B_{k+1} - B_k &=& \frac{D_k}{2} + \frac{V''}{16}\nonumber\\
4C_{k+1} - C_k &=& \frac{B_k''}{16} + \frac{3D_k''}{32} + \frac{E_k}{2}+ \frac{3F_k}{4} - \nonumber\\
&&{}-\,\frac{V'\,^2}{8}  + \frac{V^{(4)}}{512}\nonumber\\
4D_{k+1} - D_k &=& \frac{V''}{8}\\
8E_{k+1} - E_k &=& \frac{B_k''}{8} - \frac{D_k''}{8} + 3F_k + \frac{V^{(4)}}{128}\nonumber\\
16F_{k+1} - F_k &=& \frac{D_k''}{8} + \frac{V^{(4)}}{384}\ .\nonumber
\end{eqnarray}
This system is easily solved for given initial conditions. However, what we are really interested in is the continuum limit solution which is obtained by setting $k=s-1$ in the above expression and taking the limit $s\to\infty$. By doing this we are iterating our process of discretization halving from the continuum theory down to $N$. The continuum limit solution of the $p=3$ level system is simply
\begin{eqnarray}
\label{p3continuum}
V_{p=3} &=& V + \epsilon_N\frac{V''}{12} + \epsilon_N^2\left[-\frac{V'\,^2}{24}+\frac{V^{(4)}}{240}\right]\nonumber\\
(g_1)_{p=3} &=& \frac{V''}{24} + \epsilon_N\frac{V^{(4)}}{480}\\
(g_2)_{p=3} &=& \frac{V^{(4)}}{1920}\ .\nonumber
\end{eqnarray}
In the above expressions the label ``$p=3$" reminds us that this is the solution of the continuum limit of the recursive relations given in eq.~(\ref{rec:p3}) describing discretization halving at the $p=3$ level. Note that the continuum limit solution depends only on the initial potential $V$, i.e. it is not sensitive to initial values of the $g_k$'s as these terms all vanish in the continuum limit. In this way we have obtained the effective action that gives the best $N$-fold discretization of the starting theory at the $p=3$ level. One can similarly obtain a set of effective actions $S^{(p)}$, one for each value of $p$. The solution for $p=6$ is given in the Appendix. Note that each solution contains within it all the solutions for lower levels. 
Solutions for larger values of $p$ are a bit more cumbersome, however, they are just as easy to use in simulations.  Expressions up to $p=9$ can be found on our web site \cite{scl}. 

Note that one solves for the continuum limit of the level $p$ system of recursive relations but once for all theories, i.e. once the solution is found it works for all sufficiently smooth potentials $V$. Actually, the requirement for the level $p$ solution is that the starting potential is differentiable $2p-2$ times. 

The effective action satisfying the continuum limit of the discretization halving recursion relations at level $p$ leads to an $N$-fold amplitude that is equal to the continuum amplitude of the starting action up to an $O(\epsilon_N^p)$ term. Therefore, the continuum limit solution satisfies 
\begin{equation}
\label{convergence}
A^{(p)}_N(a,b;T) = A(a,b;T)+O(\epsilon_N^p)\ .
\end{equation}

Expectation values can be calculated with the same precision using standard discretized estimators, provided that the discretized time step on which those observables reside is chosen appropriately. For example, the expectation value of the momentum squared $\langle p^2(t)\rangle$ may be calculated using the standard estimator $\delta_n^2$ if the time step from $n$ to $n+1$ is shortened to $\epsilon_N^p$ keeping all the remaining time steps unchanged.

The validity of the presented analytical result will be illustrated in the following section where we present Monte Carlo simulations of two different models.
To conclude this section: we have constructed a general procedure for calculating effective actions $S^{(p)}$ for any level $p$. We have completed the procedure and found explicit values for the effective actions up to and including $p=9$. The $N$-fold amplitudes of the $p=9$ effective action differ from the continuum amplitudes by a term proportional to $1/N^9$. 
 
\section{\label{sec:numerical} Numerical results and algorithms}

\begin{figure}[!ht]
    \includegraphics[width=8.5cm]{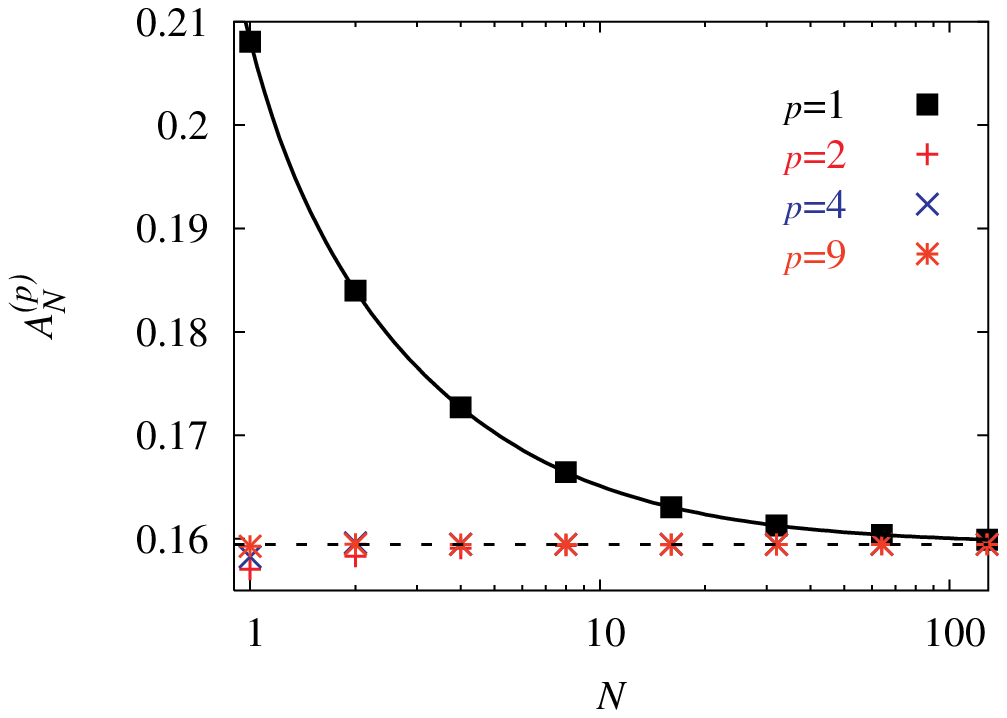}
    \includegraphics[width=8.5cm]{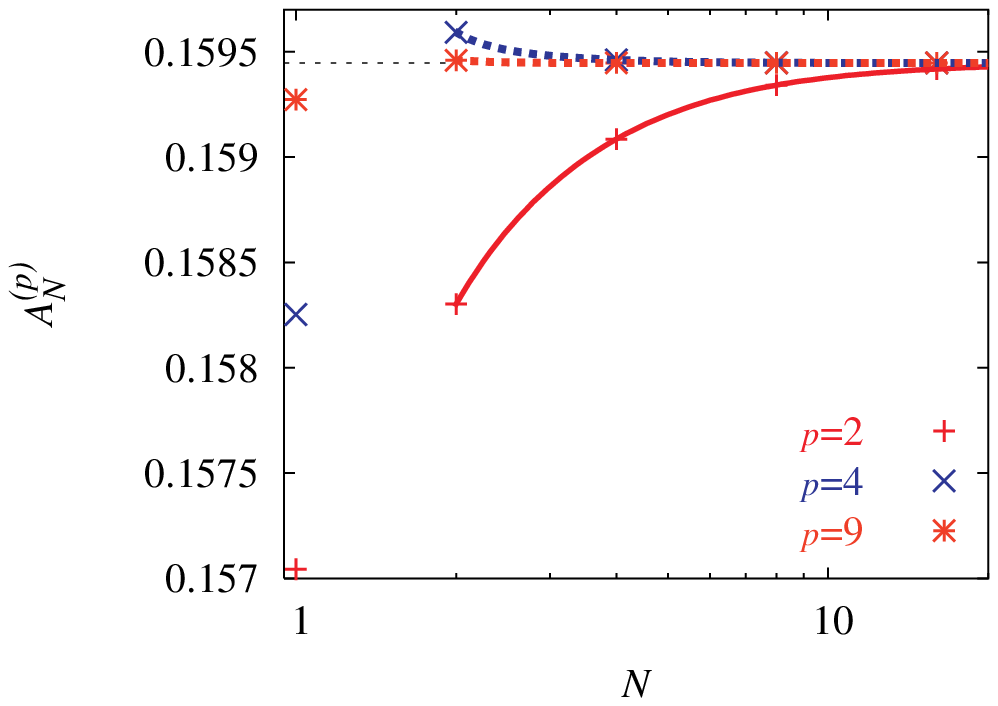}
    \caption{\label{phi4} (Color online) (top) Plot of discrete amplitudes $A_N^{(p)}$ as a function of $N$ for $p=1,2,4$ and $9$ for an 
    anharmonic oscillator with quartic coupling $\lambda=10$, time of propagation $T=1$ from $a=0$ to $b=1$, $N_{MC}=9.2\cdot 10^7$. (bottom) Detail of the same plot comparing amplitudes for $p=2,4$ and $9$. In both plots the dashed line represents the continuum limit amplitude.}
 \end{figure}

In this section we illustrate the generic results obtained in the previous sections by analyzing the speedup for different values of $p$ in the case of Monte Carlo simulations of two different models. The first model we looked at is the anharmonic oscillator with quartic coupling 
\begin{equation}
\label{anharmonic}
V(q)=\frac{1}{2}\,q^2+\frac{\lambda}{4!}\,q^4\ .
\end{equation}
In Fig.~\ref{phi4} we illustrate how the discretized amplitudes $A^{(p)}_N$ tend to the continuum limit for levels $p=1$ (naively discretized starting action), $2,4$ and $9$. The top plot gives an overall view of how the discretized amplitudes for $a=0$, $b=1$ and $T=1$ calculated using higher level effective actions systematically outperform the ones at lower level. In particular we see how they outperform the amplitude calculated using the naively discretized action. 
In the bottom plot we show a detail of the top plot which makes it easier to qualitatively track the differences between amplitudes calculated with effective actions at levels $p=2,4$ and $9$. In agreement with eq.~(\ref{convergence}), the curve fitted to the $p$ level data is a polynomial in $1/N$ of the form 
\begin{equation}
\label{fit}
A^{(p)}_N=A^{(p)}+\frac{B^{(p)}}{N^p}+\frac{C^{(p)}}{N^{p+1}}+\ldots\ . 
\end{equation}
As derived, all the $A^{(p)}$ are (within error bars) equal to each other and represent the continuum amplitude $A(a,b;T)$ we seek. The continuum value is represented in the plots by a dashed line. In all cases the fits were done for data with $N>1$. The reason that the $N=1$ points were omitted is that for $T=1$ we have $\epsilon_1=1$, i.e. these points do not satisfy the condition for asymptotic expansion $\epsilon_N<1$. However, the $N=1$ amplitudes are quite interesting because they are algebraic expressions with no integrations. From the above plot we see that the $N=1$ amplitudes calculated with higher level effective actions give better and better approximations to the amplitude. We have seen this behavior for other potentials as well.  

For $T<1$ the $N=1$ amplitudes calculated using effective actions derived in the previous section represent very good algebraic approximations for the case of a general theory -- larger levels $p$ give better approximations. From Fig.~\ref{phi4} we see that these approximations work rather well even for the marginal point $T=1$. For $T>1$ we do need to do some integrals numerically in order to get the required amplitudes. For high level effective actions it is enough to use $N=[T]+1$, i.e. to do only $[T]$ integrations numerically.    

\begin{figure}[!ht]
    \includegraphics[width=8.5cm]{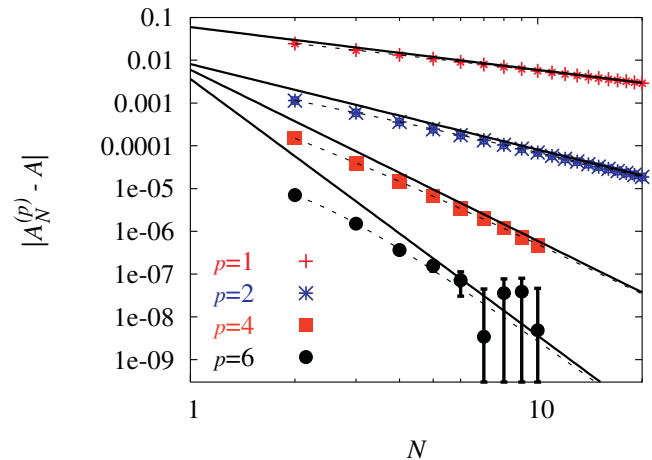}
    \caption{\label{p1246} (Color online) The deviations from the continuum limit $|A_N^{(p)}-A|$ as a function of $N$ for $p=1,2,4$ and $6$ (top to bottom). This particular plot is for the case of an anharmonic oscillator with quartic coupling $\lambda=10$, time of propagation $T=1$ from $a=0$ to $b=1$. The number of Monte Carlo samples used was $N_{MC}=9.2\cdot 10^9$ for $p=1,2$, $N_{MC}=9.2\cdot 10^{10}$ for $p=4$, and $N_{MC}=3.68\cdot 10^{11}$ for $p=6$. Dashed lines correspond to appropriate $1/N$ polynomial fits to the data. The solid lines give the leading $1/N$ behavior. The level $p$ curve has an $1/N^p$ leading behavior.}
 \end{figure}

A quantitative measure of how well the derived effective actions perform can be seen in Fig.~\ref{p1246}. We see explicitly that the $p$ level data differ from the continuum amplitudes as polynomials starting with $1/N^p$. 
Because of this, the deviations from the continuum limit $|A_N^{(p)}-A|$ become exceedingly small for larger values of $p$ making it necessary to use ever larger values of $N_{MC}$ so that the Monte Carlo statistical error does not mask these extremely small deviations. For $p=6$ we see that although we used an extremely large number of Monte Carlo samples ($N_{MC}=3.68\cdot 10^{11}$) the statistical errors become of the same order as the deviations already at $N\gtrsim 8$. For $p=9$ this is the case even for $N=2$, i.e. we already get the continuum limit within a Monte Carlo error of around $10^{-8}$.

To make the deviations in Fig.~\ref{p1246} visible for large $p$ levels we needed to run a simulation with very large $N_{MC}$. This simulation took about a week on our 160 Gflops cluster. On the other hand the simulation in Fig.~\ref{phi4} uses a much smaller $N_{MC}$ and takes less than one hour to complete. In practice we see that the derived effective actions give excellent agreement with continuum limit amplitudes already for small values of $N$. Simulations with such values of $N$ take a negligible amount of time even on a single PC. 

The second model we consider is that of a particle moving in a modified P\"oschl-Teller potential -- a well known exactly solvable model \cite{fluegge}
\begin{equation}
\label{mPT}
V(q)=-\,\frac{1}{2}\,\,\frac{\alpha^2\beta(\beta-1)}{\cosh^2 \alpha q}\ .
\end{equation}
Unlike the anharmonic oscillator the P\"oschl-Teller potential has both a continuous and discrete spectrum. The discrete eigenstates have energy
\begin{equation}
E_n=-\frac{\alpha^2}{2}(\beta-1-n)^2\ ,
\end{equation}
for $0\le n\le\beta-1$. Therefore, we see that the model has critical values of coupling $\beta=1,2,3,\ldots$ at which it acquires new bound states. 
\begin{figure}[!ht]
    \includegraphics[width=8.5cm]{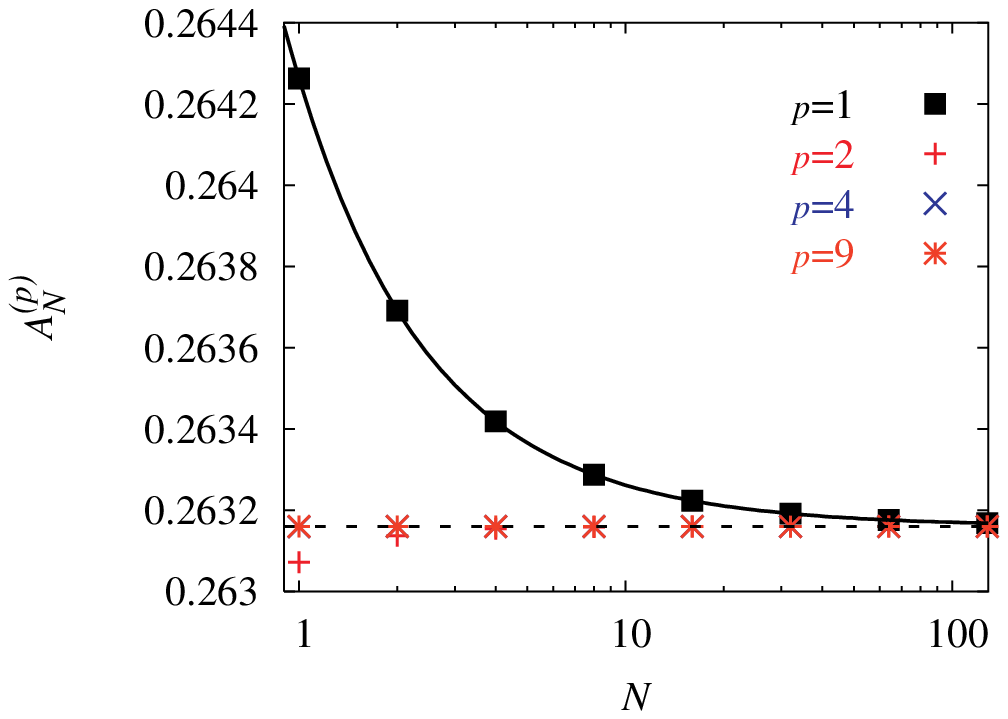}
    \includegraphics[width=8.5cm]{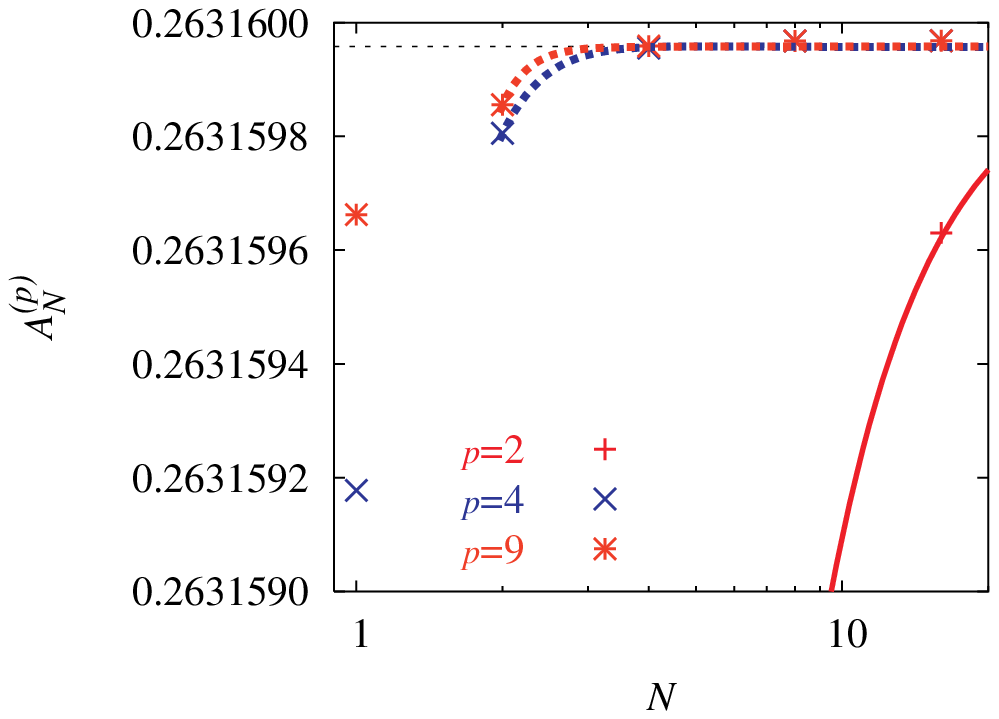}
    \caption{\label{mpt1} (Color online) (top) Plot of discrete amplitudes $A_N^{(p)}$ as a function of $N$ for $p=1,2,4$ and $9$ for a particle in a modified
    P\"oschl-Teller potential with parameters $\alpha=0.5$, $\beta=1.5$. $T=1$, $a=0$, $b=1$, $N_{MC}=9.2\cdot 10^7$. (bottom) Detail of the same plot comparing amplitudes for $p=2,4$ and $9$. In both plots the dashed line represents the continuum limit amplitude. }
 \end{figure}
\begin{figure}[!ht]
    \includegraphics[width=8.5cm]{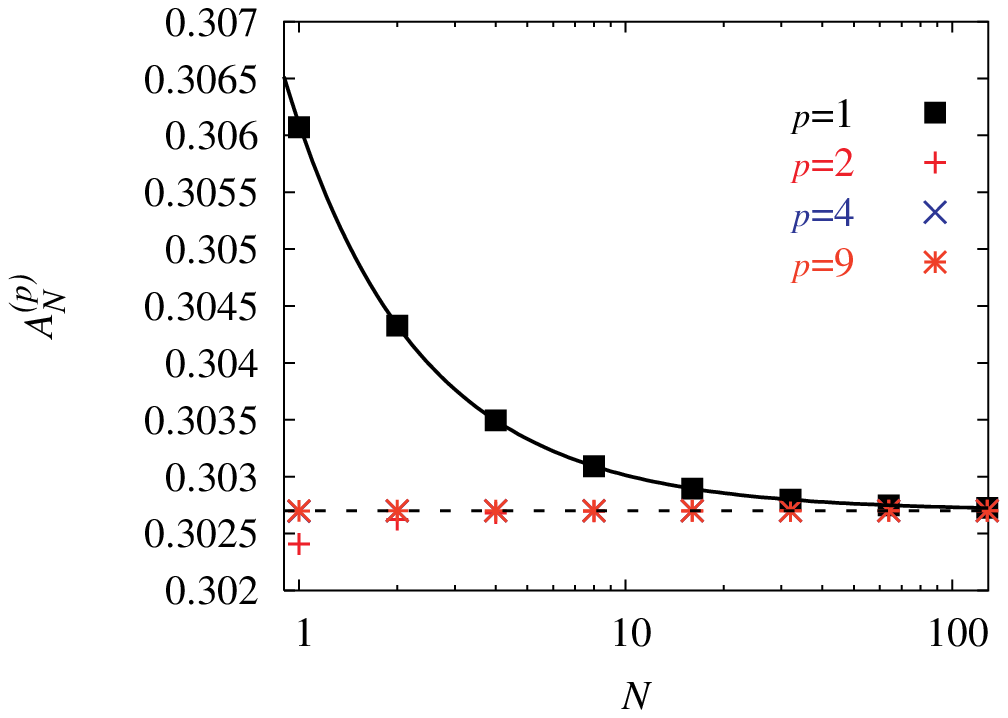}
    \includegraphics[width=8.5cm]{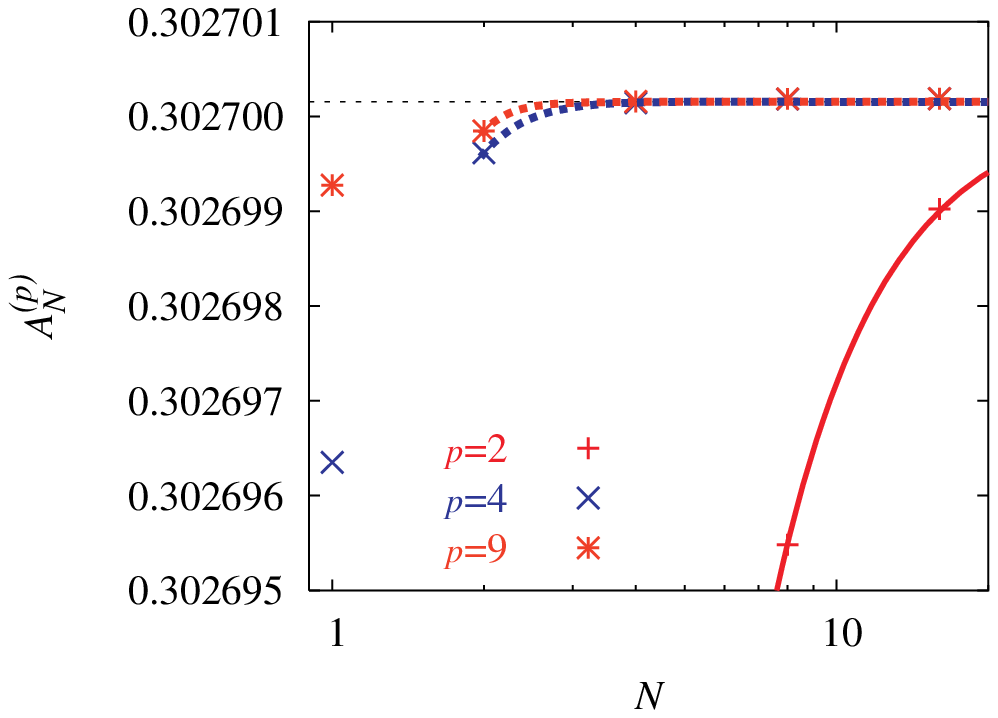}
    \caption{\label{mpt2} (Color online) Same plots as in Fig.~\ref{mpt1} but for a modified P\"oschl-Teller potential with parameters $\alpha=0.5$ and $\beta=2$.}
 \end{figure}

Fig.~\ref{mpt1} displays how the discretized amplitudes $A^{(p)}_N$ tend to the continuum limit for levels $p=1,2,4,9$ for the potential with $\alpha=0.5$ and $\beta=1.5$. The same plots for the case of $\alpha=0.5$ and $\beta=2$ (lying on the a critical value of $\beta$) are given in Fig.~~\ref{mpt2}. As we can see the effective actions work just as well as in the case of the anharmonic oscillator. Going through a critical point like $\beta=2$ certainly affects the physical quantities calculated, however, the speedup algorithm is not affected in any way.

\begin{figure}[!ht]
    \includegraphics[width=8.5cm]{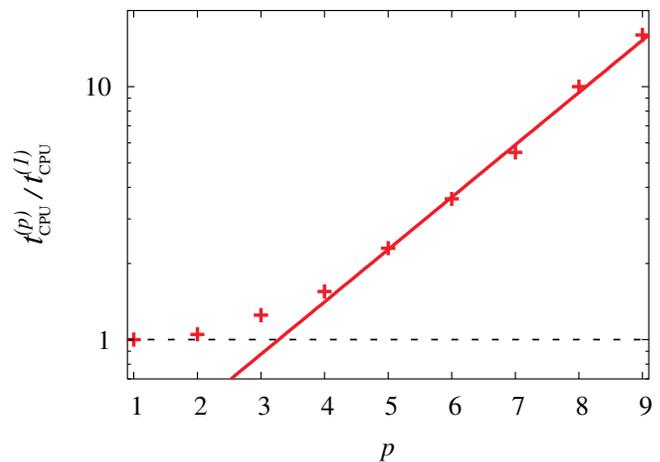}
    \caption{\label{time} (Color online) Relative increase in computation time that comes about from the increased complexity of expression for higher $p$ level effective actions.}
 \end{figure}

With the increase of $p$ level the complexity of the expressions for the effective actions grows exponentially. Therefore, the increase in computation time that results from using higher $p$ level effective actions also grows exponentially as is shown in Fig.~\ref{time}. 

As we have seen, by increasing $p$ we drastically improve convergence to the continuum limit. An important consequence of this is that we can obtain the same precission using much smaller values of $N$, i.e. much coarser discretizations. This is at the root of the speedup that we find. However, as we have seen, the exponential growth in complexity of the effective actions puts an upper bound to $p$ levels that can be used. From Fig.~\ref{time} we see that $p=9$ is still far from that upper bound -- the gain of eight orders of magnitude in the speed of convergence far outweighs what is roughly a ten fold increase in computation time. 

At the end we briefly comment on two Monte Carlo algorithms developed for simulations in this section. In the first algorithm trajectories are generated by a Gaussian distribution function obtained using a semi-classical expansion. The computing time of this algorithm scales as $N^2\cdot N_{MC}$ since it is necessary to diagonalize the quadratic form in the exponential of the distribution function. In the second algorithm we implemented the bisection method \cite{ceperley}, which scales as $N\cdot N_{MC}$. Therefore, the bisection algorithm is the method of choice for large values of $N$. On the other hand, our method allows us to obtain very precise results using small values of $N$. In that region we have found the two algorithms to be comparable both in precision and running time.  

In both algorithms we needed to use a random number generator which gives a large number of uncorrelated random numbers in a fashion suitable for parallel programming. Our primary random number generator was the Scalable Parallel Random Number Generator library \cite{srinivasanetal,sprng} (SPRNG). Following the good practise suggested by Ferrenberg et al. \cite{ferrenbergetal} we have checked all our results using a different random number generator. Checks were made with the Numerical Recipes' RAN3 generator \cite{numrec} with a different seed for each MPI process. Agreement was in all cases well within a 1-$\sigma$ interval implying that there were no hidden systematic errors present in either the algorithms or the random number generators. 

We note in passing that the analytical derivations presented in this paper work equally well in both the Euclidean and Minkowski formalism (with appropriate $i\epsilon$ regularization), i.e. they are directly applicable to quantum systems as well as to statistical ones. However, the Monte Carlo simulations used to numerically document our analytical results necessarily needed to be done in the Euclidean formalism. 

\section{\label{sec:conclusion} Conclusion}

To conclude, we have presented an algorithm that leads to significant speedup of numerical procedures for the 
calculating of path integrals of a generic theory. The increase in speed is the result of new analytical
input that has emerged from a systematic investigation of the relation between different discretizations 
of the same theory. We have presented an explicit procedure for obtaining a set of effective actions $S^{(p)}$ 
that have the same continuum limit as the starting action $S$, but which approach that limit ever faster.
Amplitudes calculated using the $N$-point discretized effective action $S_N^{(p)}$ satisfy 
$A^{(p)}_N(a,b;T)=A(a,b;T)+O(1/N^p)$, where $a$ and $b$ are initial and final states, $T$ the time of propagation,
and $A(a,b;T)$ the sought-after amplitude of the continuum theory. We have obtained and analyzed the effective actions 
for $p\le 9$. In this paper we quote expressions up to $p=6$ (see the Appendix), the rest can be found on our web site \cite{scl}. 

At the end we illustrated the obtained generic results by analyzing the speedup for different values of $p$ 
in the case of concrete Monte Carlo simulations of two different models: anharmonic oscillator with quartic coupling 
and particle in a modified P\"oschl-Teller potential. 

Extensions of the derived algorithm to $M>1$ particles and $d>1$ dimensions, as well as to quantum field theories are both in progress. In both cases 
the derivation of the analogue of integral eq.~(\ref{integral}) does not seem to present a problem. The asymptotic expansion used to solve it is also directly generalizable. However, the algebraic recursive relations that determine $S^{(p)}$ will be more complex and may practically limit us to smaller values of $p$. 

\appendix

\section{Effective Action to $p=6$}

In this Appendix we present the effective action at level $p=6$. Note that this solution contains within it the effective actions at all lower levels -- all one needs to do is to truncate the $p=6$ solution at the appropriate order in the $\epsilon_N$ expansion. For example, the effective potential $V$ at the $p=3$ level is obtained from the $p=6$ level potential by disregarding term that are $\epsilon_N^3$ and higher. Code containing $S^{(p)}$ for $p\le 9$ is available on our web site \cite{scl}.
\begin{widetext}
\begin{eqnarray}
V_{p=6}&=&V+\epsilon_N\,\frac{V''}{12}+
\epsilon_N^2\left[-\frac{V'\,^2}{24} + \frac{V^{(4)}}{240}\right]+
\epsilon_N^3\left[-\frac{V''\,^2}{360} -\frac{V'\,V^{(3)}}{120} + 
\frac{V^{(6)}}{6720}\right]+\nonumber\\
&&+\,\epsilon_N^4\left[\frac{V'\,^2\,V''}{240} - 
  \frac{23\,{V^{(3)}}^2}{40320} - 
  \frac{V''\,V^{(4)}}{1680} - 
  \frac{V'\,V^{(5)}}{2240} + \frac{V^{(8)}}{241920}\right]+\nonumber\\
&&+\,\epsilon_N^5\left[\frac{V''\,^3}{5670} + \frac{29\,V'\,V''\,
     V^{(3)}}{20160} + \frac{V'\,^2\,V^{(4)}}{2240} - 
  \frac{47\,{V^{(4)}}^2}{1209600}\,-\right.\nonumber\\ 
  &&\qquad\qquad-\left.\frac{19\,V^{(3)}\,V^{(5)}}{241920}-\frac{V''\,V^{(6)}}{30240}-\frac{V'\,V^{(7)}}{60480} + \frac{V^{(10)}}{10644480}\right]\nonumber
\end{eqnarray}
\begin{eqnarray}
(g_1)_{p=6}&=&\frac{V''}{24}+
\epsilon_N\,\frac{V^{(4)}}{480}+
\epsilon_N^2\left[-\frac{V''\,^2}{1440} - \frac{V'\,V^{(3)}}{480} + 
  \frac{V^{(6)}}{13440}\right]+\nonumber\\
&&+\,\epsilon_N^3\left[-\frac{{V^{(3)}}^2}{4032} - \frac{V''\,V^{(4)}}{5040} - 
  \frac{V'\,V^{(5)}}{6720} + \frac{V^{(8)}}{483840}\right]+\nonumber\\
&&+\,\epsilon_N^4\left[\frac{V''\,^3}{60480} + \frac{V'\,V''\,
     V^{(3)}}{3360} + \frac{V'\,^2\,V^{(4)}}{13440} - 
  \frac{13\,{V^{(4)}}^2}{806400}\,-\right.\nonumber\\ 
  &&\qquad\qquad-\left.\frac{V^{(3)}\,V^{(5)}}{26880}-\frac{V''\,V^{(6)}}{80640} - 
  \frac{V'\,V^{(7)}}{161280} + \frac{V^{(10)}}{21288960}\right]\nonumber\\
(g_2)_{p=6}&=&\frac{V^{(4)}}{1920}+
\epsilon_N\,\frac{V^{(6)}}{53760}+
\epsilon_N^2\left[-\frac{{V^{(3)}}^2}{32256} - 
  \frac{V''\,V^{(4)}}{40320} - 
  \frac{V'\,V^{(5)}}{53760} + \frac{V^{(8)}}{1935360}\right]+\nonumber\\
&&+\,\epsilon_N^3\left[-\frac{{V^{(4)}}^2}{345600} - 
  \frac{V^{(3)}\,V^{(5)}}{138240} - 
  \frac{V''\,V^{(6)}}{483840} - 
  \frac{V'\,V^{(7)}}{967680} + \frac{V^{(10)}}{85155840}\right]\nonumber\\
(g_3)_{p=6}&=&\frac{V^{(6)}}{322560}+
\epsilon_N\,\frac{V^{(8)}}{11612160}\,+\nonumber\\
&&+\,\epsilon_N^2\left[-\frac{{V^{(4)}}^2}{4147200} - 
  \frac{V^{(3)}\,V^{(5)}}{1658880} - 
  \frac{V''\,V^{(6)}}{5806080} - 
  \frac{V'\,V^{(7)}}{11612160} + 
  \frac{V^{(10)}}{510935040}\right]\nonumber\\ 
(g_4)_{p=6}&=&\frac{V^{(8)}}{92897280}+
\epsilon_N\,\frac{V^{(10)}}{4087480320}\nonumber\\
(g_5)_{p=6}&=&\frac{V^{(10)}}{40874803200}\ .\nonumber
\end{eqnarray}
\end{widetext}

\begin{acknowledgments}
The investigations presented in this paper were done at the Scientific Computing Laboratory at the
Institute of Physics in Belgrade. The authors
wish to thank the Ministry of Science and Environmental Protection of the Republic of Serbia for financing this
investigation through projects 1486 and 1899.
\end{acknowledgments}


\end{document}